\documentclass[preprint,showpacs,preprintnumbers,amsmath,amssymb,endfloats*]{revtex4}
\usepackage{graphicx}

\begin{document}
\def \ee {\varepsilon}
\thispagestyle{empty}
{\bf
Comment on ``Anomalous temperature dependence of the Casimir force
for thin metal films''
}

The Letter \cite{1} predicts ``an unusual decrease with temperature (or even
nonmonotonic temperature dependence) of the Casimir attraction force between 
a thin metal film and a bulk plane ideal metal...'' According to \cite{1},
``for bulk samples, the Casimir force {\it increases} slowly with
temperature''. On this basis the authors of \cite{1} propose the 
experimental observation 
of the decreasing temperature dependence of the Casimir force magnitude per
unit area, $|f(T)|$, in the configuration of a bulk ideal metal with 
planar boundary and a thin metal film described by the Drude model. 
As we demonstrate below, the statement of \cite{1} that for bulk samples
$|f(T)|$ increases with temperature is in error. What actually happens is
that $|f(T)|$ decreases with $T$ in a wide temperature region for bulk
samples described by the Drude model. Here, we show that this decrease is much
larger than that predicted in \cite{1} for a thin film and that it has already
been experimentally excluded. 

We have computed $|f(T)|$ for an ideal metal semispace placed at
$a=100\,$nm from a semispace made of the virtual metal considered in \cite{1}
using the Lifshitz formula. The computational results, as a function of 
temperature, are presented in Fig.~1 and should be compared with
Fig.~1(e) of \cite{1} representing respective results for a thin film
made of the same virtual metal near an ideal metal semispace. As is seen in Fig.~1,
$|f(T)|$ decreases with the increase of $T$. The comparison of both figures
shows that for two semispaces the relative decrease of $|f(T)|$ is more
pronounced than for the case of the thin film considered in \cite{1}.
Thus, for two semispaces the ratio 
$\bigl(|f(T=50K)|-|f(T)|\bigr)/|f(T=50K)|$ is equal to 1.1\%, 1.5\% and
2.2\% at $T=300\,$K, 400\,K and 600\,K, respectively. For a thin film
\cite{1} the same ratio is equal to only 0.8\%, 0.9\% and 0.5\% at the
same respective temperatures, i.e., much less than for two semispaces.
Note that for a film near a semispace the minimum value of $|f(T)|$ is
achieved at $T=400\,$K, whereas for two semispaces it is achieved at
much higher $T=2090\,$K.

According to \cite{1}, the Casimir force between two semispaces is given by
$f(T)=f_{\nu}(T)+f_{\rm rad}(T)$, where the first and second terms are
determined by the virtual and thermal fluctuations, respectively.
The Letter \cite{1} claims that at short separations $f_{\rm rad}\sim T^4$.
This is true only for two ideal metal semispaces, but is not correct
when at least one semispace is made of a metal described by the Drude model.
In the latter case $f_{\rm rad}$ is not a monotonous function of $T$ \cite{2}
and its specific $T$-dependence results in decreasing $|f(T)|$ as presented 
in Fig.~1.
Note that \cite{1} incorrectly attributes the nonmonotonous behavior of
$|f(T)|$ to the interplay between $f_{\nu}$ and $f_{\rm rad}$.
In fact for two semispaces this behavior is determined by
$f_{\rm rad}$ alone. The dependence of $f_{\nu}$ on $T$ through the
relaxation parameter does not play an important role in this effect 
and can be omitted \cite{3}.
Regarding the influence of phase transitions on the thermal Casimir force 
(metal-insulator and from normal to superconducting state), this was 
considered in detail in \cite{4,5}.

To conclude, contrary to what is claimed in \cite{1}, the observation
of the decreasing magnitude of the Casimir force with $T$ using a thin
film is disadvantageous in comparison to the case of a thick plate
(semispace). The authors of Letter \cite{1} do not inform the reader that precise
experimental determinations of the Casimir pressure between two thick
metallic plates have already been found to be inconsistent with the theoretical
description of the plate material by means of the Drude model (in \cite{6}
significant deviations between the predictions of the Drude model and
data were found, and in \cite{7,8} the Drude model was excluded at a 95\%
and 99.9\% confidence levels, respectively).
\hfill \\[3mm]
\noindent
B.~Geyer, G.~L.~Klimchitskaya,
and V.~M.~Mostepanenko\hfill \\
Institute for Theoretical
Physics, Leipzig University,
D-04009, Leipzig, Germany \hfill \\[3mm]
PACS numbers: 73.61.At, 11.10.Wx

\begin{figure}
\vspace*{-6cm}
\centerline{
\includegraphics{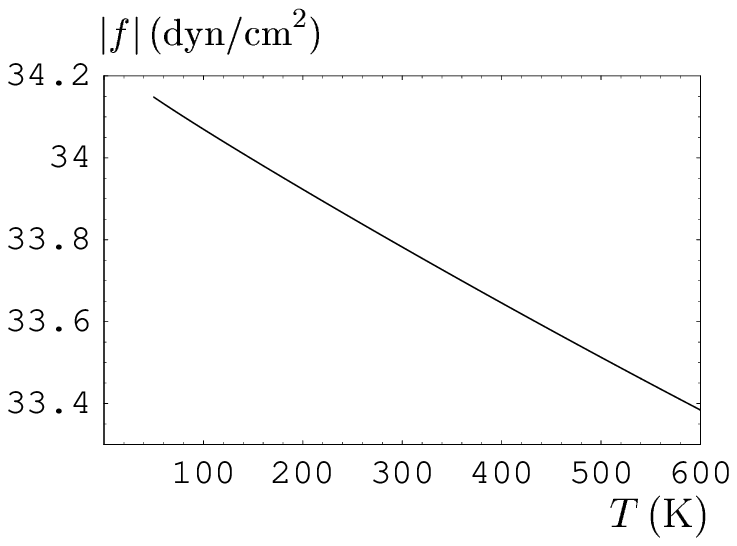}
}
\vspace*{-14cm}
\caption{The magnitude of the Casimir force per unit area vs temperature
between two semispaces made of an ideal and a virtual metal.}
\end{figure}
\end{document}